# Experimental study of the structure of a rich premixed 1,3-butadiene/CH$_4$/O$_2$/Ar flame

H.A.Gueniche, P.A. Glaude, R. Fournet, F.Battin-Leclerc

*Département de Chimie-Physique des Réactions, UMR 7630 CNRS INPL,*

*1 rue Grandville, BP 451, 54001 Nancy, France*

*glaude@ensic.inpl-nancy.fr*

**Abstract**

The structure of a laminar rich premixed 1,3-C$_4$H$_6$/CH$_4$/O$_2$/Ar flame have been investigated. 1,3-Butadiene, methane, oxygen and argon mole fractions are 0.033; 0.2073; 0.3315, and 0.4280, respectively, for an equivalent ratio of 1.80. The flame has been stabilized on a burner at a pressure of 6.7 kPa (50 Torr). The concentration profiles of stable species were measured by gas chromatography after sampling with a quartz probe. Quantified species included carbon monoxide and dioxide, methane, oxygen, hydrogen, ethane, ethylene, acetylene, propyne, allene, propene, cyclopropane, 1,3-butadiene, butenes, 1-butyne, vinylacetylene, diacetylene, C$_5$ compounds, benzene, and toluene. The temperature was measured thanks to a thermocouple in PtRh (6%)-PtRh (30%) settled inside the enclosure and ranged from 900 K close to the burner up to 2100 K.

**Introduction**

Soots and polyaromatic hydrocarbons (PAH) present in the exhaust gas of engines and combustion devices represent a high part of urban pollution. Their formation in combustion involves small unsaturated hydrocarbons for which the chemistry is still very uncertain. Different ways are proposed for the formation and the oxidation of the first aromatic compounds, involving the reactions of C$_2$ (acetylene), C$_3$ or C$_4$ unsaturated species [1-5].

Allene and propyne flames have been previously investigated [6] to study the determinant role of the propargyl radicals in benzene formation. Among the other unsaturated soot precursors implied in the aromatic ring formation, 1,3-butadiene is of primary importance as a primary product in hydrocarbon oxidation and because of its structure involving two conjugated unsaturated bonds. Furthermore, 1,3-butadiene is an interesting molecule in studying the formation of benzene, due to the possibility of branching between C$_2$, C$_3$, and C$_4$ chains. The oxidation of butadiene has been already studied in some oxidation conditions: shock tube [7], flow reactor [8], and premixed flat flame [1]. The previous work in flame involved a mixture oxygen/1,3-butadiene/argon 67.5/29.5/3.0 corresponding to an equivalence ratio of 2.4 under a pressure of 2.67 kPa (20 Torr), in near-sooting conditions.

The purpose of the present work is to experimentally investigate the structure of a premixed flat 1,3-C$_4$H$_6$/CH$_4$/O$_2$/Ar flame. These results will be use in a further extend to improve the mechanism previously developed in our laboratory for the reactions of C$_3$-C$_4$ unsaturated hydrocarbons (propyne, allene, 1-butyne, 2-butyne, 1,3-butadiene) [7,9].

**Experimental procedure**

The experiments were performed using an apparatus newly developed in our laboratory to study temperature and stable species profiles in a laminar premixed flat flame at low pressure. The body of the flat flame matrix burner, provided by McKenna Products, was made of stainless steel, with an outer diameter of 120 mm and a height of 60 mm (without gas/water connectors). The burner, used in this investigation, was built with a bronze disk (95% copper, 5% tin). The porous plate (60 mm diameter) for flame stabilization was water cooled (water temperature: 60°C) with a cooling coil sintered into the plate. The burner could be operated with an annular co-flow of argon to favor the stabilization of the flame.

This burner was housed in a water-cooled vacuum chamber evacuated by two primary pumps and maintained at 50 Torr (6.7 kPa) by a regulation valve. This chamber was equipped of four quartz windows for an optical access, a probe for samples taking and a thermocouple for temperature measurements. The burner could be vertically translated, while the housing and its equipments were kept fixed. A sighting telescope measured the position of the burner relative to the probe or the thermocouple with an accuracy of 0.01 mm. The flame was lighted on using an electrical discharge.

Gas flow rates were regulated by RDM 280 Alphagaz and Bronkhorst (El-Flow) mass flow regulators. 1,3-Butadiene had a 99.5 % purity, methane a 99.95 % purity and both were supplied by Alphagaz - L'Air Liquide. Oxygen (99.5% pure) and argon (99.995% pure) were supplied by Messer.

Temperature profiles were obtained using a PtRh (6%)-PtRh (30%) type B thermocouple (diameter 100 µm). The thermocouple wire is sustained horizontally by a fork and crosses the flame to avoid conduction. The junction is put at the centre of the burner. In the case of the flame perturbed by the probe, the junction of the thermocouple is located at two probe's inlet diameter or so from the probe's tip, corresponding to about 200 µm. The thermocouple was coated with an inert layer of $BeO-Y_2O_3$ to prevent catalytic effects [10]. The ceramic was done by damping in a hot solution of $Y_2(CO_3)_3$ (93% mass.) and BeO (7% mass.) followed by a drying in a Meker burner flame. This process was reiterated until the whole metal was covered. Radiative heat losses were corrected using the electric compensation method [11]. The temperature measured by the thermocouple had been first calibrated as a function of the heating electrical power in a high vacuum. The real temperature of the gas was obtained thereafter at the intersection of the calibration curve and the experimental one measured in the flame.

Stable species profiles were determined by gas chromatography. Gas samples were directly obtained by connecting the quartz micro probe to a volume, which was previously evacuated by a turbo molecular pump down to $10^{-7}$ kPa and which was then filled up to a pressure of 10 Torr (1.3 kPa). The diameter of the extremity of the probe was about 100 µm, while the upper part diameter was 6 mm, with a angle of ~40°. The sudden decrease of the pressure from the flame gases to the inside of the probe insured all reactions to be frozen.

Chromatographs with a Carbosphere packed column and helium or argon as carrier gas were used to analyse $O_2$, $H_2$, CO and $CO_2$ by thermal conductivity detection and $CH_4$, $C_2H_2$, $C_2H_4$, $C_2H_6$ by FID. Heavier hydrocarbons (a-$C_3H_4$, p-$C_3H_4$, c-$C_3H_6$, $C_3H_6$, $C_4H_2$, $C_4H_4$, 1,3-$C_4H_6$, 1-$C_4H_6$, 2-$C_4H_8$, $C_5$ compounds, benzene, and toluene) were analysed on a haysep packed column by flame ionisation detection (FID) and nitrogen as gas carrier gas. Water was not directly measured, but its mole fraction was determined by mass balance on O atom. The change in the mole fractions due to the dilution by the reaction products was taken into account from a first simulation. Some small oxygenated compounds were detected but were not quantitatively analysed.

**Results and Discussion**

A methane flame doped with 1,3-butadiene have been investigated. The gas flow rate was 3.32 l/min and the mixtures contained 20.7% methane, 3.3% 1,3-butadiene, 33.2% oxygen and 42.8% argon corresponding to an equivalent ratio of 1.80. $C_4$ compounds represented then 16% of the amount of methane. Figure 1 presents the temperature profiles obtained for the flame. The lower temperatures were measured at 0.04 cm from the burner and were around 600 K without the probe, 440 K with the probe. The higher temperatures were reached between 0.7 and 0.9 cm above the burner and were around 2150 K without the probe. The temperature decreased thereafter because of the heat losses.

Figure 2 to 11 present the profiles of the reactants and of the products of the reaction in function of the height above the burner. Figure 2 displays the profiles of the fuels, methane and 1,3-butadiene, and figure 3 those of oxygen, hydrogen and water that was calculated from the mass balance. Carbon oxides are presented on figure 4. Figure 5 shows the profiles of $C_2$ products, figures 6 and 7 those of $C_3$ products, figures 8 and 9 those of $C_4$ products. Some $C_5$ products, such as methyl butadiene or cyclopentadiene, have been measured. They have not been displayed here since they have not been calibrated yet. Figure 10 displays the profile of benzene and figure 11 that of toluene that is the heaviest analyzed compound.

The consumption of both fuels occurs in the first stage of the flame, but 1,3-butadiene is consumed a little bit closer to the burner, at 5 mm height while some methane remains up to 7 mm. The higher reactivity of the butadiene can be explained by the weaker C-H bonds on the tertiary C-atoms of this reactant (99.9 kcal/mol [9]) that allows easier H-abstractions yielding a resonance stabilized i-$C_4H_5$ ($CH_2$=CH—(•)C=$CH_2$) radical:

$$C_4H_6 + R\bullet \rightarrow RH + i\text{-}C_4H_5$$

Furthermore the two double bonds involved in this fuel allow the addition of small radicals, especially H atoms yielding a resonance stabilized butenyl free radical.

$$C_4H_6 + H \rightarrow C_4H_7$$

The shape of the profile of oxygen is very similar to that of methane (figure 3). Final carbon-containing products are CO and $CO_2$ (figure 4). Since the mixtures are rather rich, φ = 1.80, CO remains the major carbon oxide in the post-flame. Hydrogen is also one of the major products with a mole fraction reaching 0.2 in the end gas. This species is produced by the H-abstractions by H-atom on the fuels. The low amount of oxygenated free radicals caused by the high equivalence ratio does not permit to oxidize thereafter $H_2$ to water.

The main stable intermediates are $C_2$ molecules, which are mainly yielded by the combustion of methane (figure 5). Ethane and ethylene are produced promptly and reach their higher amount close to the burner. In this flame, the peak of ethylene is located a little bit closer to the burner than that of ethane, while that of acetylene is located further at 0.6 cm. These molecules are produced in the combustion of methane via the combination of methyl radicals that is yielded by a H-abstraction from the reactant.

$$CH_3 + CH_3 \rightarrow C_2H_6$$

The adduct can also decompose directly

$$CH_3 + CH_3 \rightarrow C_2H_5 + H$$

Ethane reacts by H-abstraction and yields also $C_2H_5$ radicals, that decompose quickly into ethylene

$$C_2H_5 \rightarrow C_2H_4 + H$$

The prompt formation of ethylene, which peak is reached before that of ethane, can be due to reaction of 1,3-butadiene, for instance by addition with H-atoms followed by the decomposition of the adduct:

$$C_4H_6 + H \rightarrow C_2H_4 + C_2H_3$$

A minor amount of ethylene react by addition with small radicals, but the main channel is the H-abstraction by H or OH yielding $C_2H_3$. These radicals oxidize then with $O_2$ or decompose into acetylene.

$$C_2H_3 \rightarrow C_2H_2 + H$$

Since this last molecule is the most stable in flame conditions, its amount remains higher than that of other $C_2$ species. In the rich mixtures studied here, the amount of acetylene decreases slowly in the post flame.

Among the $C_3$ products, the most important is propene. It can come from both oxidation of methane and butadiene. $C_3H_6$ can be produced by the termination of methyl radicals with $C_2H_3$ or its addition on $C_2H_4$ followed by the decomposition of the adduct

$$CH_3 + C_2H_4 \rightarrow C_3H_6 + H$$

$C_3H_6$ can also be produced by addition of methyl radicals on butadiene followed by the decomposition of the adduct:

$$CH_3 + C_4H_6 \rightarrow C_3H_6 + C_2H_3$$

In this case, it is mostly resulting of a coupling between the two fuels since methyl radical are yielded by a H-abstraction on methane.

Cyclopropane must be mainly formed by the molecular isomerization of propene.

Propyne and allene are also present among the stable intermediates in the flame (figure 7). They can be produced by many pathways involving $C_2$, $C_3$ and $C_4$ compounds. A first route yielding propyne is the addition of methyl radicals on acetylene followed by the decomposition of the adduct:

$$C_2H_2 + CH_3 \rightarrow p\text{-}C_3H_4 + H$$

The H-abstractions on propene produced allyl radicals that decompose also to allene or propyne. The H-abstractions on 1,3-butadiene yield resonance stabilized $i\text{-}C_4H_5$ that can decompose to allene and a methyl radical. The amounts of allene and propyne are linked together by the molecular isomerization that is easier in the high temperature conditions.

The major $C_4$ products are 1-butyne, 2-butene (figure 8), vinylacetylene and diacetylene (figure 9), which are produced in the very first stage of the combustion. Even if some routes from the $C_2$ species are involved in their formation, they must be yielded mainly in the combustion of 1,3-butadiene. 2-Butene can be produced by the termination reaction between resonance stabilized $i\text{-}C_4H_5$ and H-atom while 1-butyne can be produced by successive molecular isomerizations. 1,3-butadiene isomerized to 1,2-butadiene that produced then 1-butyne. 1,2-Butadiene has not been quantified experimentally since it could not be separated from the reactant 1,3-butadiene by chromatography.

Vinylacetylene is produced directly from 1,3-butadiene by a H-abstraction yielding $i\text{-}C_4H_5$ followed by a decomposition by β-scission:

$$C_4H_6 + R\bullet \rightarrow RH + i\text{-}C_4H_5$$

$$C_4H_5 \rightarrow C_4H_4 + H$$

The same scheme explains the formation of diacetylene from vinylacetylene:

$$C_4H_4 + R\bullet \rightarrow RH + C_4H_3$$

$$C_4H_3 \rightarrow C_4H_2 + H$$

The heaviest and most unsaturated compounds quantified here are benzene (figure 10) and toluene (figure 11). In these conditions, benzene can be produced by both the $C_3$ and $C_4$-pathways, as proposed by many authors [2-4]. A first route is the combination of propargyl radicals that are produced by allene and propyne:

$$C_3H_3 + C_3H_3 \rightarrow C_6H_6$$

$$C_3H_3 + C_3H_3 \rightarrow C_6H_5 + H$$

The adduct can stabilized or decompose into phenyl radical and H-atom.

Another channel involves the $C_4$ compounds deriving from 1,3-butadiene. $C_4H_5$ radical can add on acetylene yielding benzene plus H:

$$C_4H_5 + C_2H_2 \rightarrow C_6H_6 + H$$

$C_2H_3$ can add too on $C_4H_4$, the adduct react then by internal addition to form a benzene ring:

$$C_2H_3 + C_4H_4 \rightarrow C_6H_6 + H$$

Toluene can be produced by the addition of methyl on benzene followed by the elimination of a H atom:

$$C_6H_6 + CH_3 \rightarrow C_7H_8 + H$$

It can be produced too by the termination step between phenyl and methyl radicals:

$$C_6H_5 + CH_3 \rightarrow C_7H_8$$

Some addition reaction $C_5$ and $C_2$ compounds followed by internal rearrangements could also be involved in the formation of toluene.

**Conclusion**

This paper presents new experimental results for a rich premixed laminar flat flame of methane and 1,3-butadiene. Profiles of temperature and of stable species have been measured for many products and in peculiar many unsaturated compounds that are soot precursors. The first aromatic rings, benzene and toluene have been measured and in these conditions, their formation seems due to both $C_3$ and $C_4$ routes.

Future work will include a kinetic modeling of these data to improve the reaction mechanisms of small soot precursors responsible for the first aromatic rings formation. Flames doped with other unsaturated compounds such as cyclopentene will be also investigated.

**Acknowledgements**

Financial support of this work by the European Union within the "SAFEKINEX" Project EVG1-CT-2002-00072 is gratefully acknowledged.

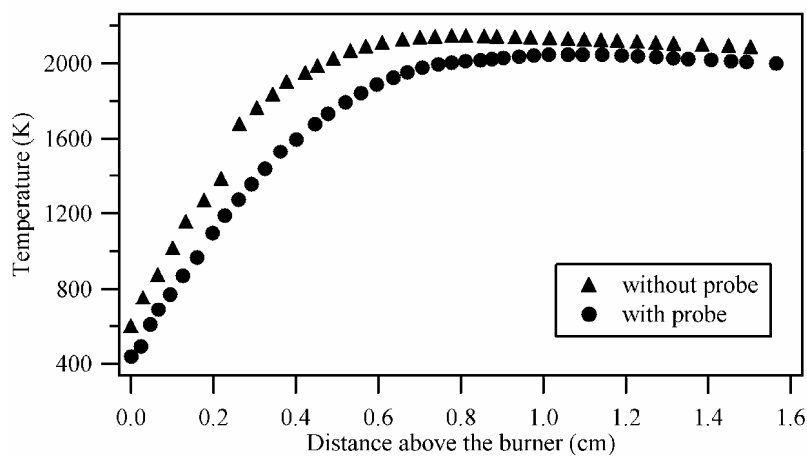

*Figure 1: Experimental temperature profiles for the 1,3-butadiene flames with and without probe perturbation, measured with a coated type B thermocouple.*

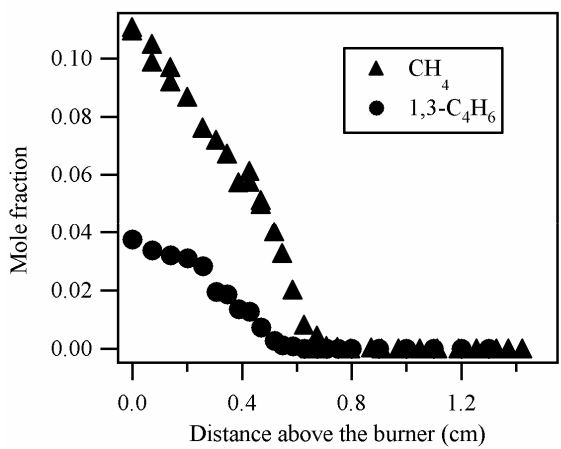

*Figure 2: Mole fraction of the fuels.*

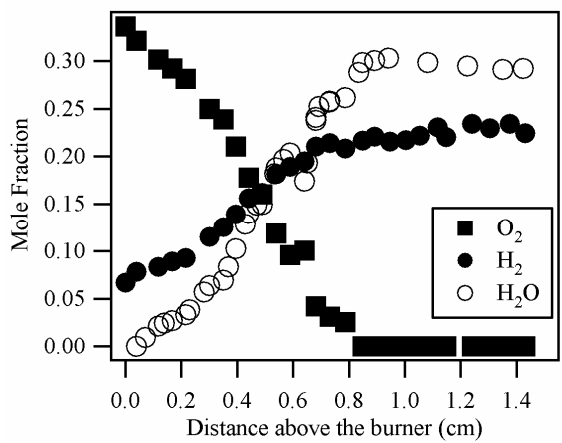

*Figure 3: Mole fraction of oxygen, hydrogen and water*

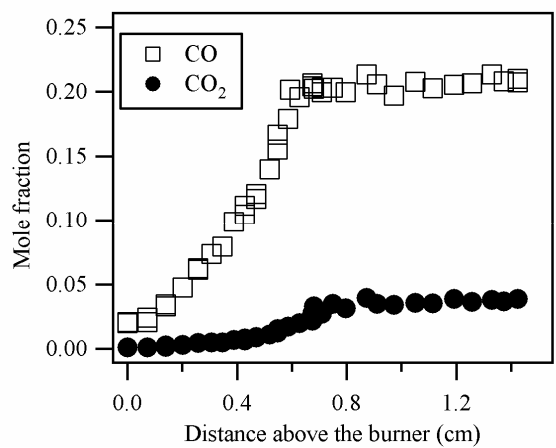

*Figure 4: Mole fraction of carbon oxides.*

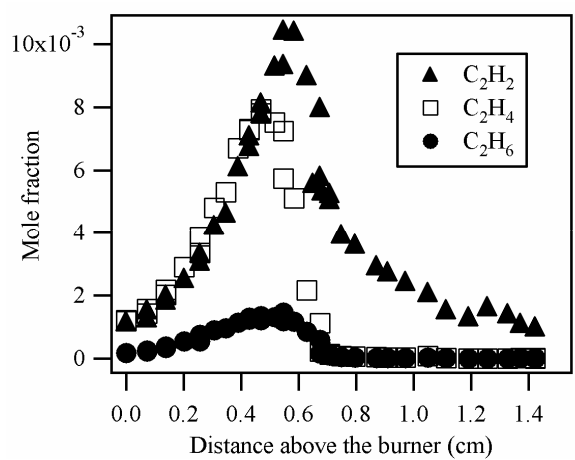

*Figure 5: Mole fraction of ethane, ethylene and acetylene.*

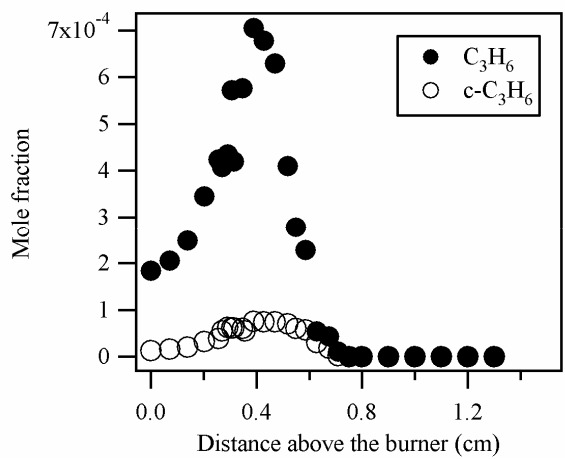

*Figure 6: Mole fraction of propane and cyclopropane.*

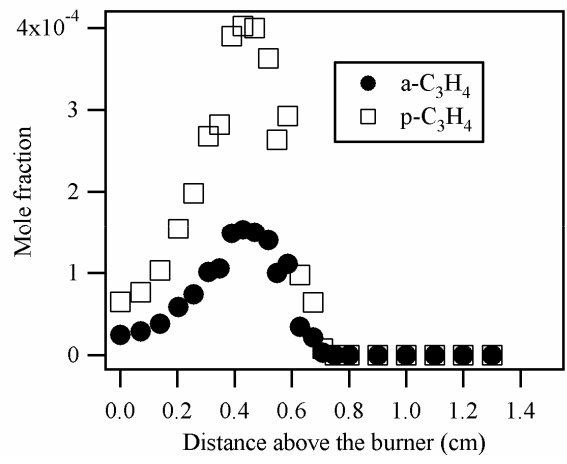

*Figure 7: Mole fraction of allene (propadiene) and propyne.*

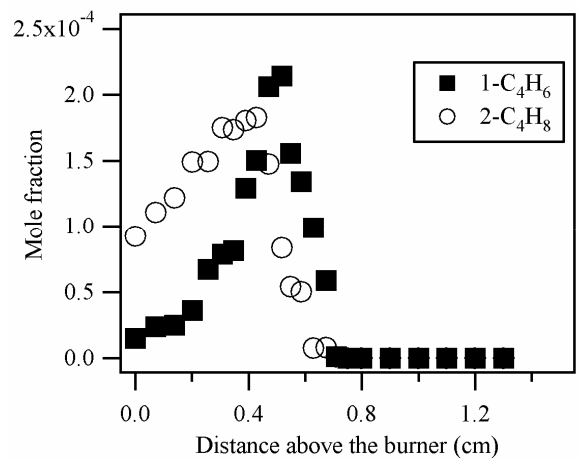

*Figure 8: Mole fraction of 1-butyne and 2-butene.*

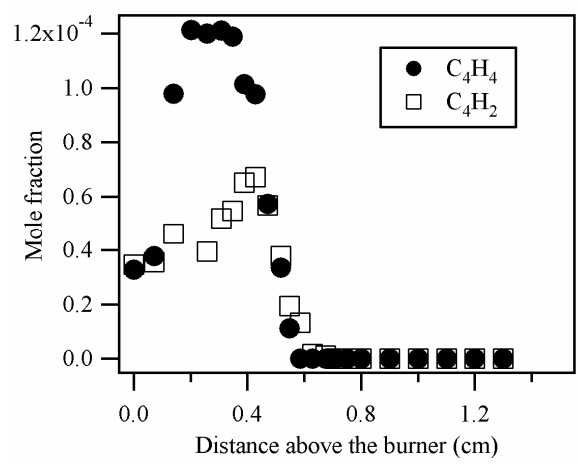

*Figure 9: Mole fraction of vinylacetylene and diacetylene.*

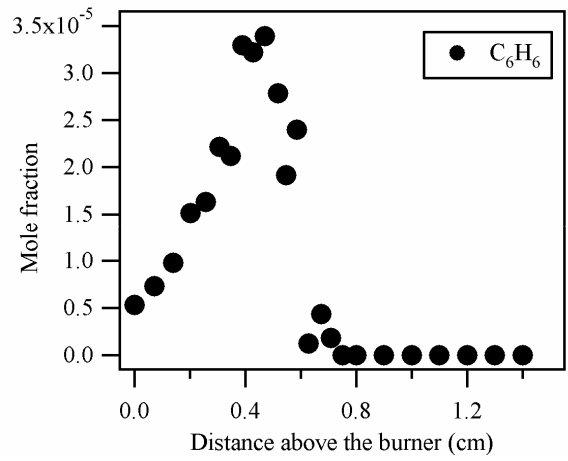

*Figure 10: Mole fraction of benzene.*

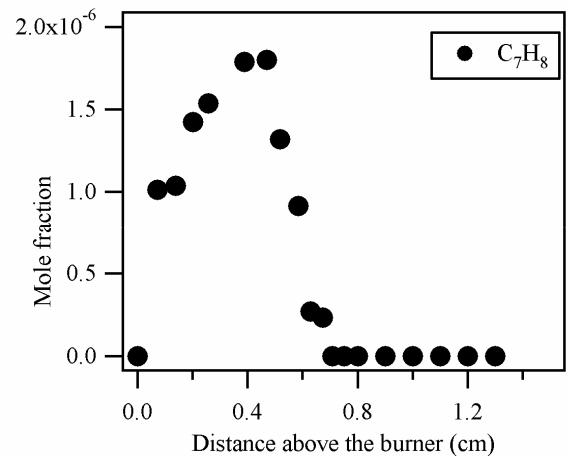

*Figure 11: Mole fraction of toluene.*